\documentclass[3p,times]{elsarticle}
\usepackage{ecrc}
\usepackage{amsmath}
\volume{00}
\firstpage{1}
\journalname{Nuclear Physics A}
\runauth{A. Ficnar, J. Noronha and M. Gyulassy}
\jid{nupha}
\usepackage{amssymb}
\usepackage[figuresright]{rotating}

\begin{document}

\begin{frontmatter}
\title{Non-conformal Holography of Heavy Quark Quenching}
\author{Andrej Ficnar, Jorge Noronha and Miklos Gyulassy}
\address{Department of Physics, Columbia University, 538 West 120th Street, New York, NY 10027, USA}
\begin{abstract}
We develop a holographic (bottom-up) gravity model for QCD to understand the connection between the peak in the trace anomaly and the magnitude of heavy quark energy loss in a strongly-coupled plasma. The potential of the scalar field on the gravity side is constructed to reproduce some properties of QCD at finite temperature and its parameters are constrained by fitting lattice gauge theory results. The energy loss of heavy quarks (as predicted by the holographic model) is found to be strongly sensitive to the medium properties. \end{abstract}
\end{frontmatter}

\section{Introduction}

One of the most important properties of the quark-gluon plasma created in heavy ion collisions at the Relativistic Heavy Ion Collider is its near-ideal fluid behavior, indicated by the small ratio of shear viscosity to entropy density $\eta/s$ \cite{luzum-romatschke}, which we assume to imply strong coupling. A very useful tool to study strongly coupled gauge theories is the AdS/CFT correspondence: in its original form \cite{maldacena}, AdS/CFT correspondence is an equivalence of a (3+1)-dimensional $\mathcal{N}=4$ super-Yang-Mills (SYM) gauge theory and the type IIB string theory on $AdS_5\times S^5$ spacetime. Using this conjecture and by taking the limit $N_c\gg \lambda \gg 1$ one can study this gauge theory at strong coupling by studying classical, two-derivative (super)gravity. 

Because of this very useful aspect of the AdS/CFT correspondence, one is tempted to use thermal $\mathcal{N}=4$ SYM to study and understand properties of finite-temperature QCD. However, there are significant differences between these two theories, most of them which originate from the conformal invariance of $\mathcal{N}=4$ SYM: this gives rise to a trivial, temperature-independent speed of sound ($c_s=1/\sqrt{3}$), vanishing bulk viscosity, absence of phase transition, etc. In addition, lattice gauge theory results for the QCD trace anomaly \cite{bnl-columbia} show significant violation of the conformal invariance near the crossover temperature $T_c$. Therefore, in order to capture these important phenomenological features of finite-temperature QCD, we are led to consider conformally non-invariant gravity theories with thermodynamics properties that can match the relevant QCD phenomenology.

\section{Constructing the potential}

According to the holographic dictionary of the original AdS/CFT correspondence, a scalar field (dilaton) $\phi$ in the bulk of $AdS_5$ space is dual to the ${\rm Tr}F^2$ operator in the $\mathcal{N}=4$ SYM at the boundary of that space. Assuming that this dictionary holds in general and that the thermodynamics in large-$N_c$ QCD is dominated by the adjoint degrees of freedom, we will look for models with a non-trivial dilaton profile, with aim to reproduce the non-trivial QCD thermodynamics. The simplest bottom-up effective realization of a gravity theory dual to a non-conformal gauge theory is a 5-dimensional gravity theory coupled to a scalar field (dilaton):
\begin{equation}\label{pot1}
S=\frac{1}{2\kappa_5^2}\int d^5 x\sqrt{-G}\left(R-\frac{1}{2}(\partial\phi)^2-V(\phi)\right)
\end{equation}
where $V(\phi)$ is the dilaton potential, $\kappa_5$ is the 5-dimensional gravitational constant and $G_{\mu\nu}$ is the 5-dimensional metric.

The first constraint will come from the requirement that the potential should give asymptotically ($\phi\to0$) $AdS_5$ spacetime, which translates into the conformal invariance of the dual field theory in UV. In order to obtain the $AdS_5$ spacetime with radius $L$ as $\phi\to 0$, the potential must asymptote to a negative constant, so in general it needs to have the following form:
\begin{equation}\label{pot2}
V(\phi)=-\frac{12}{L^2}+\frac{1}{2}m^2\phi^2+\mathcal{O}(\phi^4)
\end{equation}
Such a potential translates into a relevant deformation of the conformal field theory: $\mathcal{L}_{CFT}\to\mathcal{L}_{CFT}+\Lambda_\phi^{4-\Delta}\mathcal{O}_\phi$. Here $\Lambda_\phi$ is the scale of the deformation and $\Delta$ is the UV dimension of the field theory operator $\mathcal{O}_\phi$ dual to $\phi$. The dimension of the operator is determined from $\Delta(\Delta-4)=m^2L^2$ with $\Delta\ge2$ (the larger root) and $\Delta<4$ in order for the deformation to be relevant. This requirement of asymptotically conformal field theory originates from the idea \cite{gubser-prl} of making the connection with QCD by matching the dimension $\Delta$ of $\mathcal{O}_\phi$ to the dimension of ${\rm Tr}F^2$ in QCD at some UV scale $Q$; this UV matching to QCD essentially means that the asymptotic freedom of QCD gets replaced (or approximated) by conformal invariance. Therefore, our dual theory will not be asymptotically free, which is also already visible from the fact that $\eta/s$ is $1/(4\pi)$ at all temperatures \cite{Kovtun:2004de}, since we neglected the higher derivative terms in the action (\ref{pot1}). This in turn means that the validity of our model should not be extended to temperatures too far above or too far below the crossover temperature $T_c$.

A natural choice for one of the terms in the potential is $\cosh(\gamma\phi)$, since this term interpolates between a constant potential for small $\phi$ (the CFT limit) and $e^{\gamma\phi}$ (which gives a constant, $\gamma$-dependent speed of sound) for large $\phi$ and, as discussed in \cite{gubser-nellore}, should give a temperature-dependent speed of sound. The coefficient $\gamma$ gives the value of the speed of sound in the IR (and will be numerically determined by fitting the lattice results), but it also fixes the dimension $\Delta$, according to (\ref{pot2}). In order to have more freedom in choosing this dimension, we will introduce an additional $\phi^2$ term in the potential. We will also introduce $\phi^4$ and $\phi^6$ terms to fine-tune the fit of the cross-over behavior of the lattice results. Finally, our potential has the following form \cite{gubser-prl,gubser-nellore}:
\begin{equation}\label{pot4}
V(\phi)=\frac{1}{L^2}\left(-12\cosh(\gamma\phi)+b_2\phi^2+b_4\phi^4+b_6\phi^6\right).
\end{equation}
A similar bottom-up approach has been studied in \cite{kiritsis-q,kajantie}, where the authors explored the role played by the gauge theory's beta function in the construction of the scalar field potential in the gravity theory.

\section{Fitting the lattice}
We will use the following ansatz for the metric in the Einstein frame
\begin{equation}\label{lat1}
ds^2=G_{\mu\nu}dx^\mu dx^\nu=e^{2A(r)}\left(-h(r)dt^2+d\vec{x}^2\right)+e^{2B(r)}\frac{dr^2}{h(r)},\;\;\phi=\phi(r),
\end{equation}
which is dictated by the required symmetries \cite{gubser-nellore} and where the SO(3,1) boost invariance in $(t,\vec{x})=(t,x_1,x_2,x_3)$ direction is broken by the finite temperature, i.e. presence of the black hole with regular horizon at some finite $r=r_H$ defined by $h(r_h)=0$. In this ansatz we still have some gauge freedom left to reparametrize the radial direction $r$, so we will follow \cite{gubser-nellore} and use the gauge choice $\phi(r)\equiv r$, which means that the conformal (UV) boundary is at $r=0$. With this gauge choice and ansatz (\ref{lat1}) we can solve the equations of motion from the action (\ref{pot1}) with potential of the form (\ref{pot4}) using the method developed in \cite{gubser-nellore} in which one constructs a nonlinear 'master' differential equation for a generating function $G(\phi)$, from which one then obtains all three unknown functions in (\ref{lat1}), $A(\phi)$, $B(\phi)$ and $h(\phi)$. 

Once we solve the equations of motion, we can find the entropy density and temperature using Hawking's formulas:
\begin{equation}\label{lat2}
s=\frac{2\pi}{\kappa_5^2}e^{3A(r_H)},\;\;T=\frac{1}{4\pi}e^{A(r_H)-B(r_H)}\left|h'(r_H)\right|\,.
\end{equation}
From here we can then find the speed of sound as $c_s^2=d\log T/d\log s$ and other thermodynamic quantities. Since we now have the speed of sound $c_s^2(T)$ dependent on the parameters of the potential (\ref{pot4}), we can numerically determine these parameters by fitting our results to the lattice QCD data (with dynamical fermions) from the hotQCD collaboration \cite{bnl-columbia}. Satisfactory fits were obtained by using the following values for the parameters in the potential: $\gamma= 0.606,\;\; b_2=0.703,\;\; b_4=-0.12,\;\; b_6=0.00325$, which corresponds to dimension $\Delta\approx 3$. Our fit together with the lattice results is shown in Figure 1 (leftmost plot) and after fitting the speed of sound all other thermodynamic quantities are automatically fitted as well (e.g. trace anomaly, center plot). In the plots also indicated is the critical temperature $T_c$, which we define as the temperature where the speed of sound has a minimum. One must also define a rule for converting the temperature in GeV (as a function of which the lattice results are reported) and the temperature in $1/L$ units, or the 'AdS' units (as a function of which one obtains holographic results), in order to be able to compare both of them on the same plot. We do that by finding $T_{min}$ in AdS units, where the holographic speed of sound reaches minimum, and by postulating a linear conversion rule $T(GeV)/T_c=T(AdS)/T_{min}$, we obtain the same shapes of the thermodynamical quantities as functions of temperature as the ones from the lattice.
\begin{figure}[h!]
	  \centering
    \includegraphics[width=54mm]{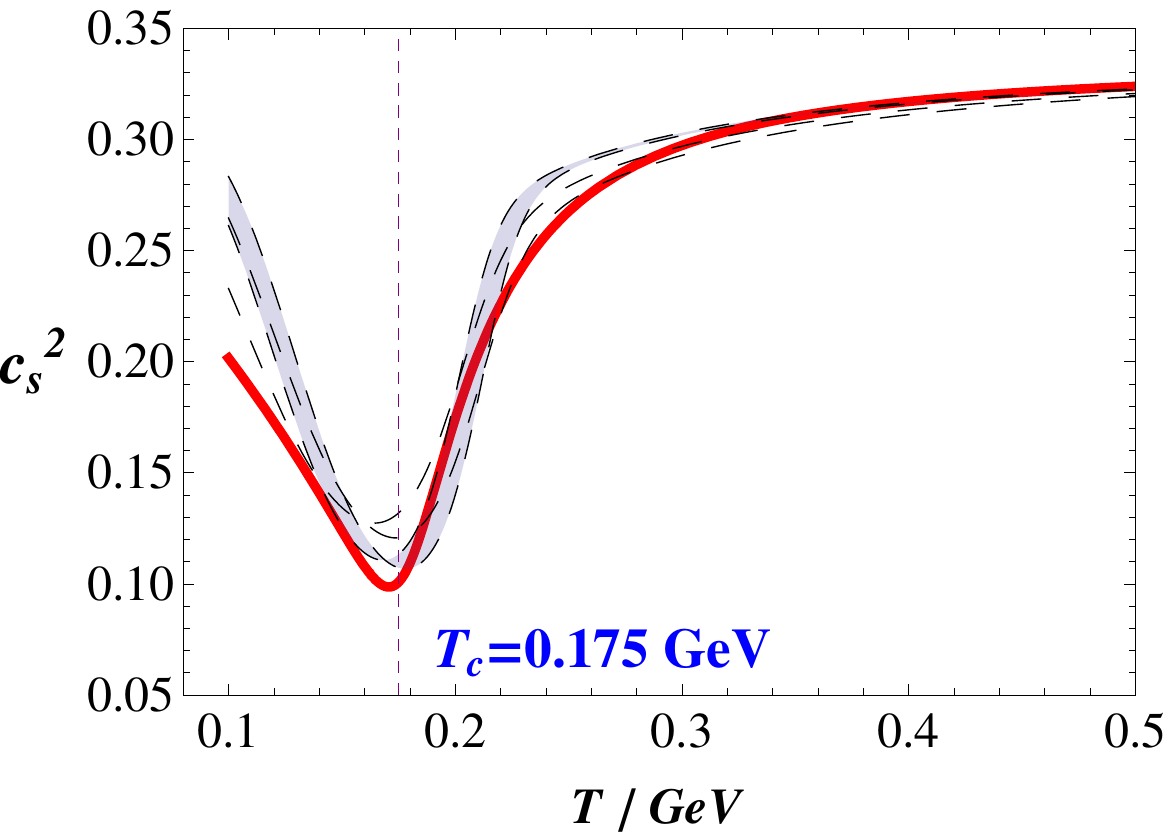}
    \includegraphics[width=54mm]{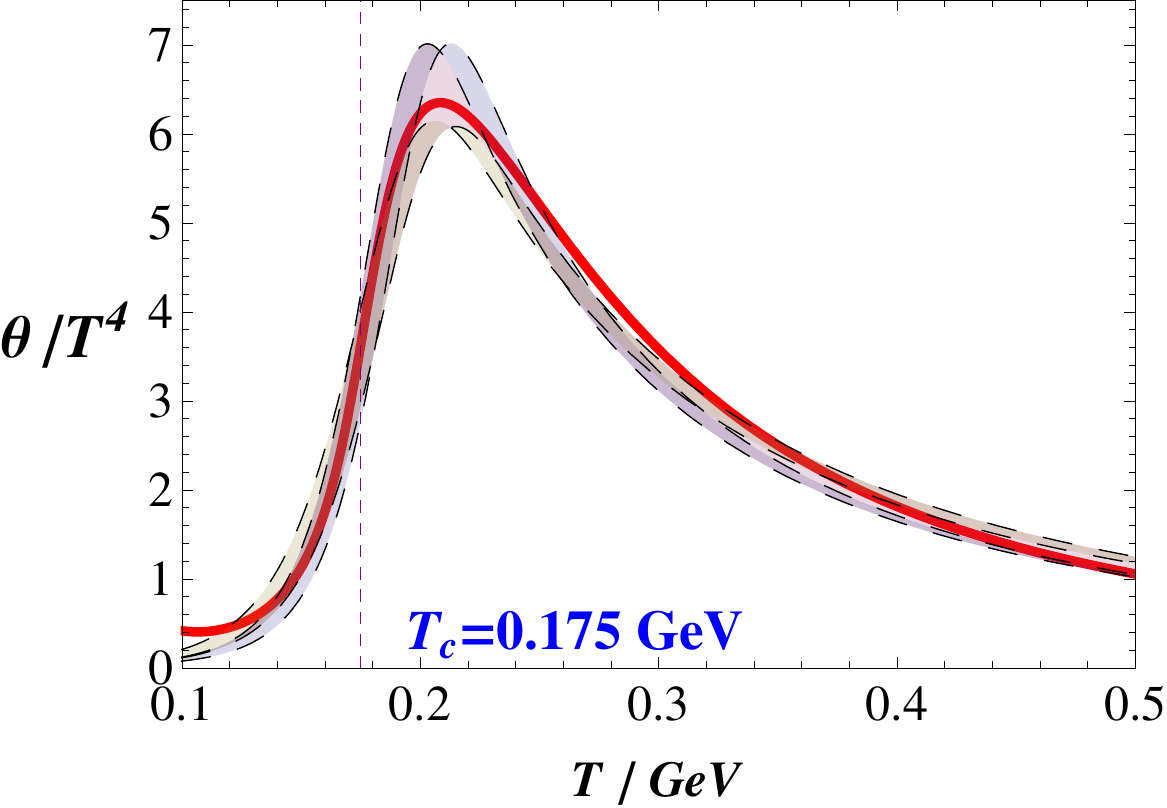}
    \includegraphics[width=54mm]{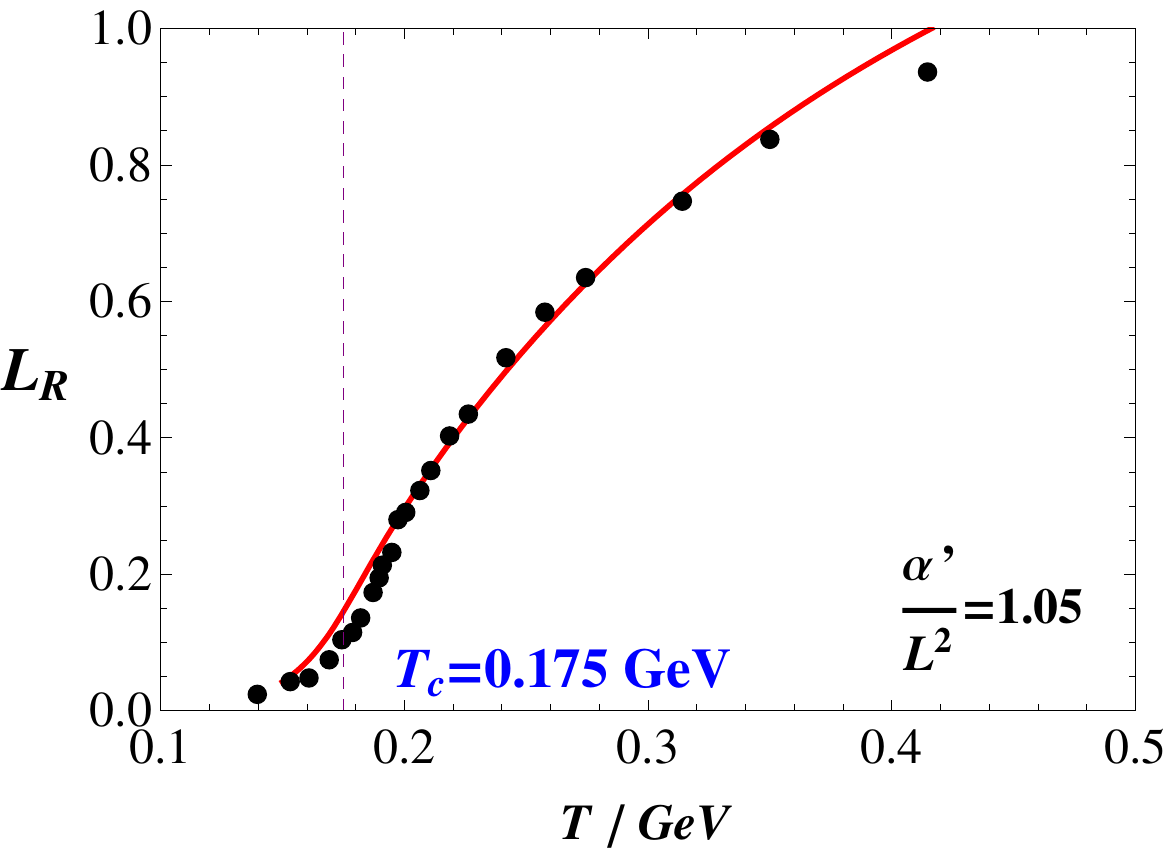}\\
    \parbox[l]{162mm}{\footnotesize{\textbf{Figure 1.}} Comparison of the speed of sound (left), trace anomaly $\theta=\epsilon-3p$ (center) and the renormalized Polyakov loop (right) as a function of temperature from our non-conformal holographic model (red curves) and the lattice results from the hotQCD collaboration \cite{bnl-columbia} (dashed curves and points). Also indicated is the critical temperature $T_c$ which we define as the temperature where the speed of sound reaches minimum.}
\end{figure}

\section{Heavy quark energy loss}
We now introduce a quark of mass $m_Q$, which is dual to a curved string in the bulk stretching from a D4 brane at $r_m\sim~1/m_Q$ to the horizon $r_H$ \cite{karch-katz}. We are working in the probe approximation, where one neglects the backreaction of the metric due to the introduction of the string. Assuming quantum corrections to be negligible, the dynamics of the string is governed by the classical Nambu-Goto action:
\begin{equation}\label{eloss1}
S_{NG}(\mathcal{D})=\frac{1}{2\pi\alpha'}\int\limits_\mathcal{D}d^2\sigma q(\phi)\sqrt{\det(G_{\mu\nu}\partial^aX^\mu\partial^bX^\nu)}
\end{equation}
where $\alpha'=l_s^2$, squared fundamental string length, $\sigma^a$ are the string worldsheet coordinates, $X^\mu$ are its spacetime coordinates and $q(\phi)$ is the coupling between the string and the dilaton $\phi$, which originates when one expresses the string frame metric (which directly enters the Nambu-Goto action) in terms of the Einstein frame metric (\ref{lat1}). For this coupling we use the result from the five-dimensional non-critical string theory $q(\phi)=e^{\sqrt{\frac{4}{3}}\phi}$ \cite{kiritsis-q}. According to the gauge/gravity dictionary, Polyakov loops are related to the string action:
\begin{equation}\label{eloss3}
\left|\left< L(T) \right>\right|=e^{-F_Q(T)/T}\sim e^{-S_{NG}(\mathcal{D})}
\end{equation}
and since the Nambu-Goto action (\ref{eloss1}) contains $\alpha'$ (at this moment the only undetermined parameter in our model), we can use the lattice results for the Polyakov loop \cite{bnl-columbia} and by fitting the prediction of our model to the data (using the procedure from \cite{jorge-loops}), determine $\alpha'$. In that way, our model is completely constrained by the lattice data and the energy loss then becomes a prediction.
From the fit in the rightmost plot in Figure 1, we see that we get roughly $L^2/\alpha'\sim 1$. This means that, in principle, $\alpha'$-corrections (in the string worldsheet fluctuations and in the effective gravity action) can be important in this case, and they must be explicitly calculated in order to see their effect on thermodynamics and on the energy loss.

Assuming that the quark is heavy enough and that the range of relevant temperatures is low enough so that $r_m\ll r_H$, we can evaluate the quark energy loss by using the trailing string model \cite{drag-force}:
\begin{equation}\label{eloss4}
\frac{dE}{dx}=-\frac{v}{2\pi\alpha'}e^{2A(r_*)}q(\phi(r_*))
\end{equation}
where $v$ is the velocity of the quark (assumed uniformly moving) and $r_*$ is defined as $h(r_*)=v^2$. Of course, in the CFT limit this expression becomes the familiar drag force $\sim \sqrt{\lambda}T^2\gamma v$ \cite{drag-force}. 

Assuming that the trailing string model is applicable to both charm and bottom quarks (in the sense that $r_m\ll r_H$) and that they obey the usual relativistic dispersion relation $p=\gamma m v$, we can compute their energy loss in our non-conformal holographic model using the formula (\ref{eloss4}). Figure 2 shows the energy loss as a function of temperature for a fixed momentum and as a function of momentum for several fixed temperatures. In all plots the  energy loss from the non-conformal model is compared to the CFT limit of our model, where the same $L^2/\alpha'\sim 1$ was used. 
\begin{figure}[h!]
	  \centering
	  \includegraphics[width=54mm]{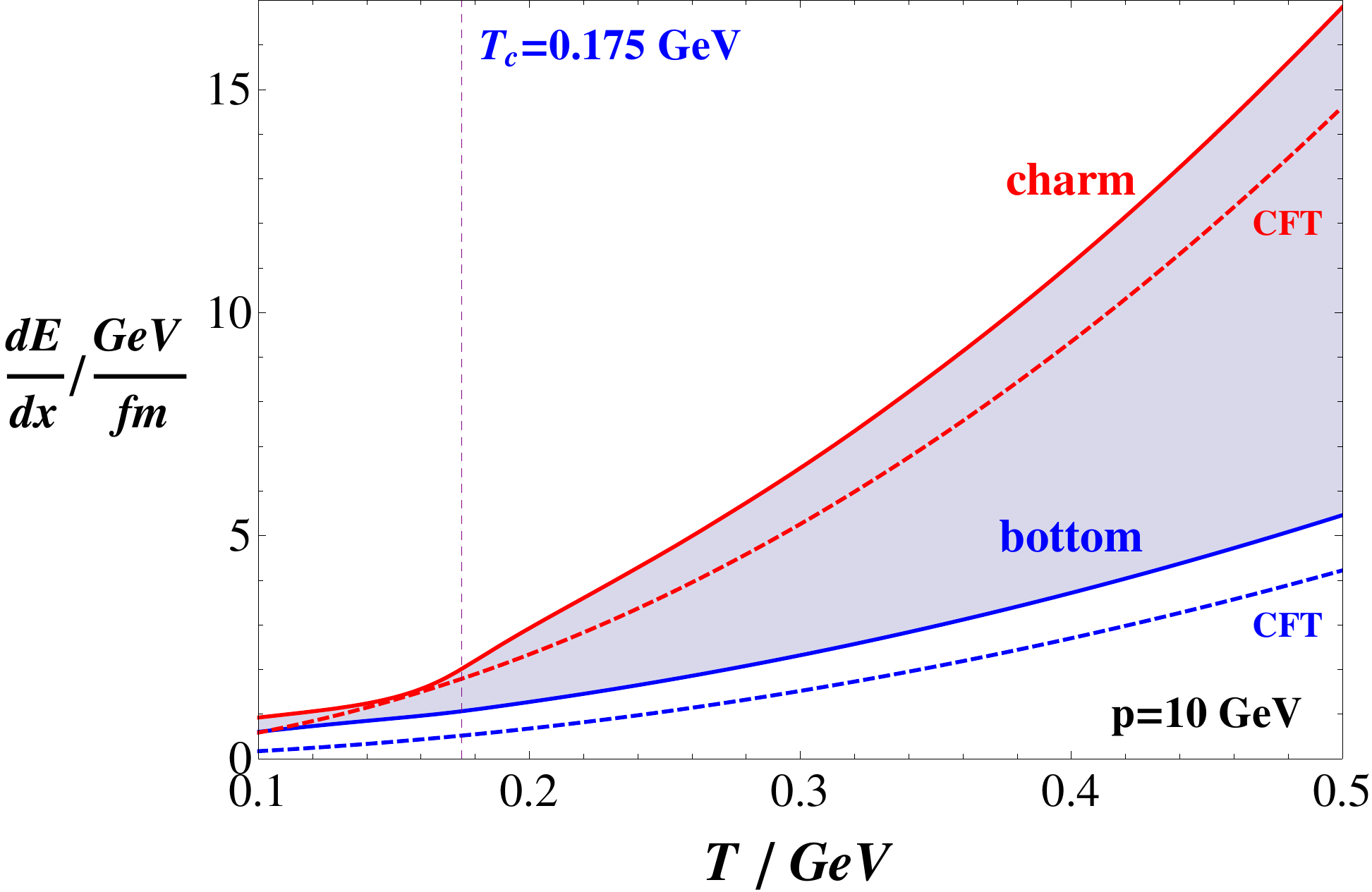}
	  \includegraphics[width=54mm]{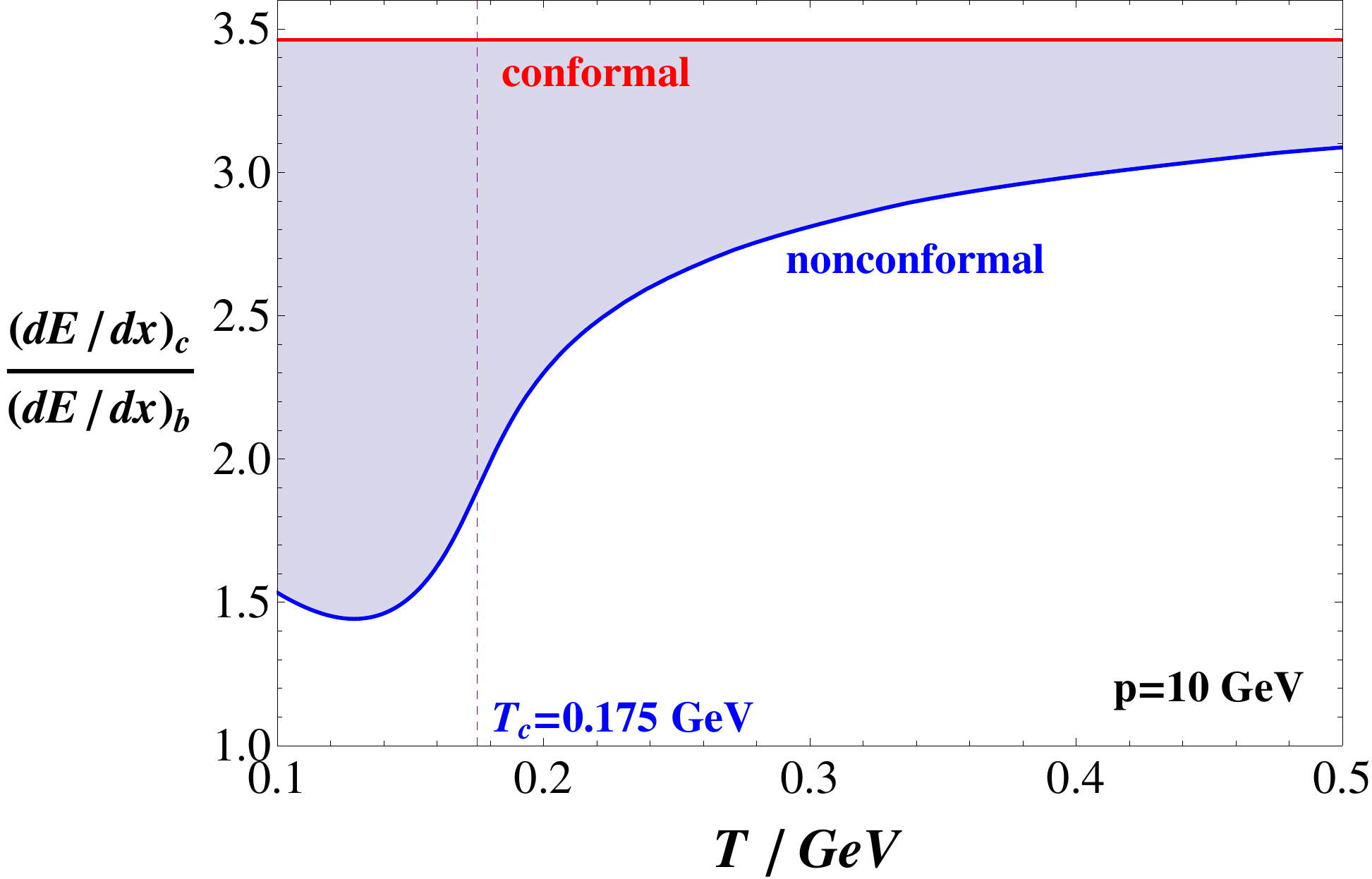}
	  \includegraphics[width=54mm]{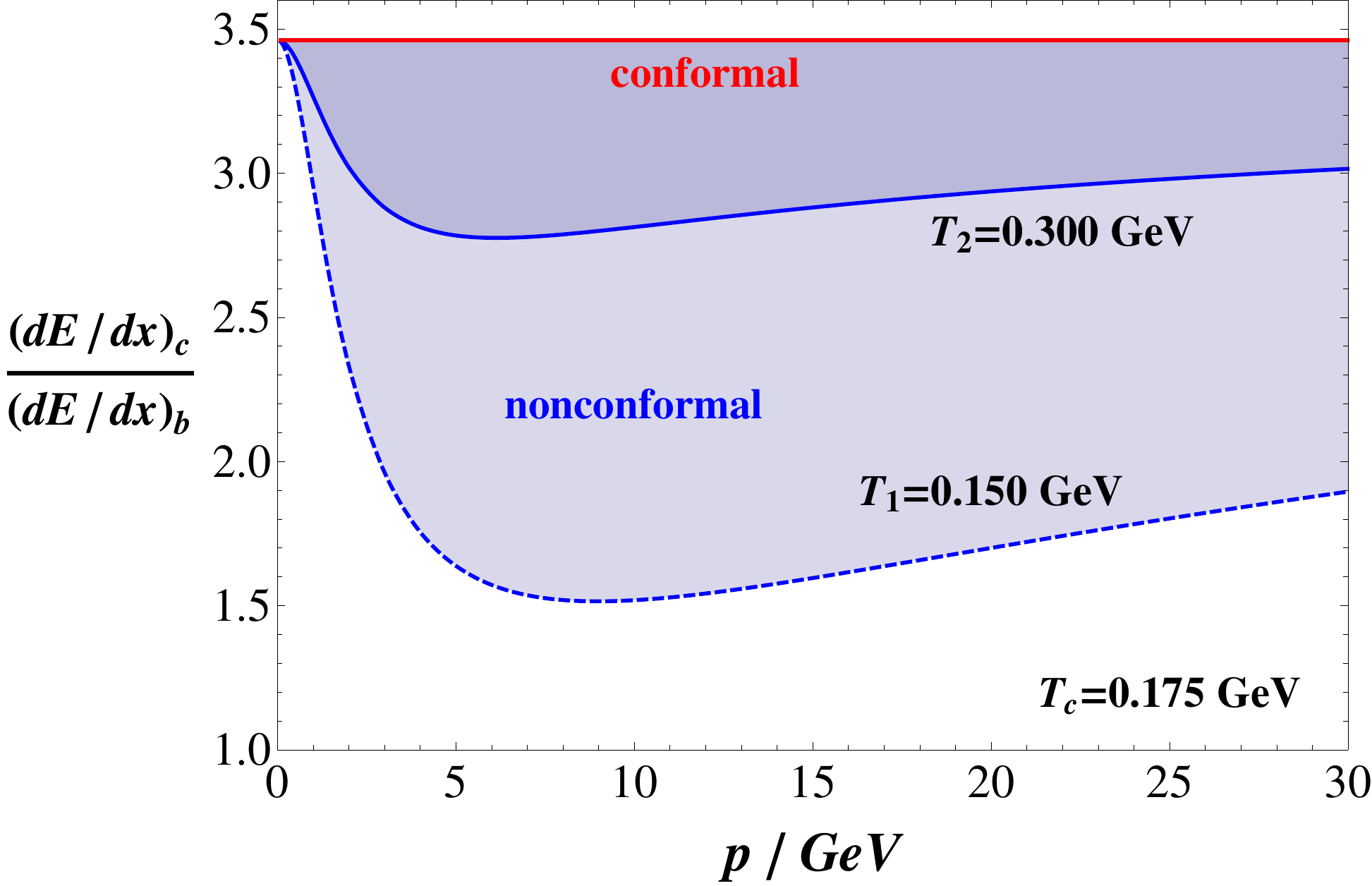}\\
   \parbox[l]{162mm}{\footnotesize{\textbf{Figure 2.}} Energy loss for charm and bottom quarks (and their ratios) as a function of the temperature for fixed momentum $p=10$ GeV (left and center) and as a function of momentum for several fixed temperatures (right) predicted in our non-conformal holographic model, compared to the CFT limit of the model.}
\end{figure}

From Figure 2 we see that a significant deviation from the CFT results is found for temperatures around $T_c$, where the trace anomaly is maximal. This difference should affect the relative ratio of the energy loss observables, such as $R_{AA}$, since we expect that a quark traveling through a realistic, expanding quark-gluon plasma spends significant amount of time in the temperature region around $T_c$.

\section{Conclusions and outlook}
Gauge/string duality gives us means to address the violation of conformal invariance in a strongly coupled plasma near $T_c$ and see how the trace anomaly and the details of equation of state affect the energy loss of heavy quarks. We showed that, in principle, the details of equation of state are important for the heavy quark energy loss in the crossover region (near $T_c$) and may affect observables such as the nuclear modification factor of jets in heavy ion collisions \cite{our-paper}. 

We thank W. Zajc and A.\ Dumitru for helpful discussions. We acknowledge support by US-DOE Nuclear Science Grant No. DE-FG02-93ER40764.

\bibliographystyle{elsarticle-num}

\end{document}